# Interrelation of fast and slow electron waves at propagation of electromagnetic waves in Maxwellian collisionless plasma[1]

V. N. Soshnikov [2]

Plasma Physics Dept.,

All-Russian Institute of Scientific and Technical Information

of the Russian Academy of Sciences

*(VINITI, Usievitcha 20, 125315 Moscow, Russia)*

*It is shown in linear approximation that in the case of one-dimensional problem of transverse electron waves in a half-infinite slab of homogeneous Maxwellian collisionless plasma with the boundary field frequency ω two wave branches of solution of the dispersion equation are simultaneously realizing. These are the branch of fast forward waves determined mainly by Maxwell equations of electromagnetic field, as well as the branch of forward and backward slow waves determined in the whole by kinetic properties of electrons in the collective electrical field. The physical nature of wave movements is revealed. A relation is found between electric field amplitudes of fast and slow waves. Multiform dividing the coupled slow waves into standing and traveling parts leads to a necessity of additional requirements to a selection of the type of a device analyzing these waves and its response interpretation.*

Solving the half-infinite homogeneous Maxwellian plasma slab problem of transverse electron waves with boundary field $E = E_0 \exp i\omega t$ and zero initial conditions using Laplace transform leads to asymptotical solution for Laplace components (images) $E_{p_1 p_2}$ (the electric field) and $f_{p_1 p_2}$ (the perturbed part of electron distribution function) [1], [2]:

$$E_{p_1 p_2} G_{p_1 p_2} = p_2 E_{p_1} + F_{p_1} - \frac{4\pi e n_e p_1}{c^2} \int \frac{v_x v_z f_{p_1}}{p_1 + v_x p_2} d\vec{v}, \tag{1}$$

$$f_{p_1 p_2} = \left( \frac{1}{p_1 + v_x p_2} \right) \left[ v_x f_{p_1} + \frac{e}{m} \frac{\partial f_0}{\partial v_z} \frac{1}{G_{p_1 p_2}} \left( p_2 E_{p_1} + F_{p_1} - \frac{4\pi e n_e p_1}{c^2} \int \frac{v_x v_z f_{p_1}}{p_1 + v_x p_2} d\vec{v} \right) \right], \tag{2}$$

where $x$ is direction of wave propagation; $e \equiv |e|$ is positive value of electron charge; $E_{p_1}$, $f_{p_1}$, $F_{p_1}$ are Laplacian images of functions $E(0, t)$, $f(\vec{v}, 0, t)$, $\left. \dfrac{\partial E(x, t)}{\partial x} \right|_{x=0}$,

$$f_0 = \left( \frac{m}{2\pi k_B T} \right)^{3/2} e^{-\frac{mv^2}{2k_B T}}, \tag{3}$$

$$G_{p_1 p_2} = p_2^2 - \frac{p_1^2}{c^2} - \frac{\omega_L^2}{c^2} p_1 \frac{m}{k_B T} \int \frac{v_z^2 f_0}{p_1 + v_x p_2} d\vec{v} \approx p_2^2 - \frac{p_1^2}{c^2} \left( 1 + \frac{\omega_L^2}{p_1^2 - \overline{v_x^2} p_2^2} \right); \tag{4}$$

---

[2] Krasnodarskaya str., 51-2-168, Moscow 109559, Russia. E-mail: vikt3363@yandex.ru .

$p_1 \equiv i\omega$ (for time component), and $p_2 \equiv ik$ (for coordinate component); $m$ is electron mass; $T$ is electron temperature; $c$ is light velocity in vacuum; $\omega_L$ is Langmuire frequency $\left(\dfrac{4\pi e^2 n_e}{m}\right)^{1/2}$; $f_{p_1}$ is function of velocity $v_x$; $F_{p_1}$, $E_{p_1}$ are independent of $\vec{v}$.

Proceeding from an approximate relation (4) and following further, one used evaluations of the type of described in works [1], [2] with the principal value sense indefinite integrals (cf. the grounds in [3]). In this way we have four roots of dispersion equation $G_{p_1 p_2} = 0$ at $p_1 = i\omega$

$$p_2^{(1\pm)} \approx \pm i \frac{\omega}{c}\sqrt{1 - \omega_L^2/\omega^2} \qquad (5)$$

for background and forward fast waves with the wave length $\lambda_{fast} \approx \dfrac{2\pi c}{\omega\sqrt{1 - \omega_L^2/\omega^2}}$, and for backward and forward slow waves with the wave length $\lambda_{slow} \approx 2\pi \dfrac{\sqrt{\overline{v_x^2}}}{\omega} \ll \lambda_{fast}$

$$p_2^{(2\pm)} \approx \pm \frac{i\omega}{\sqrt{\overline{v_x^2}}}\left(1 + \frac{\overline{v_x^2}\,\omega_L^2}{2c^2\omega^2}\right) \approx \pm i\frac{\omega}{\sqrt{\overline{v_x^2}}}. \qquad (6)$$

Owing to linearity of the problem one passes to real value solution with picking out real value parts of the boundary field and of asymptotical solutions for $E(0,t)$ and $f_1(\vec{v},x,t)$. At a given boundary field $E(0,t)$ for the considered self-consistent problem the other boundary conditions for distribution function are not arbitrary and must entirely be determined with the given field $E(0,t)$.

According to the classic method of Laplace transform, asymptotic solution of the problem appears as a sum of exponents $A_{lm}\exp\left(p_1^{(l)}t + p_2^{(m)}x\right)$ with amplitudes $A_{lm}$ which are determined by the residua of functions $f_{p_1 p_2}$ and $E_{p_1 p_2}$ in the pair-poles of $G_{p_1 p_2} = 0$.

Peculiar additional solution is here asymptotic solution of equation (2) corresponding to the pair-poles $p_1 = i\omega$, $p_2 = -i\omega/v_x$ of Laplacian image $f_{p_1 p_2}$ with $G_{p_1 p_2} \neq 0$. This solution corresponds to the so called kinematic wave $\exp\left(i\omega t - i\omega x/v_x\right)$ with an amplitude determined by the residuum

$$W_{kinem} = \left\{ (p_1 - i\omega)\left[ v_x f_{p_1} + \frac{e}{m}\frac{\partial f_0}{\partial v_x}\frac{1}{G_{p_1 p_2}}\left( p_2 E_{p_1} + F_{p_1} - \frac{4\pi e n_e p_1}{c^2}\int \frac{v_x v_z f_{p_1}}{p_1 + v_x p_2}\,d\vec{v}\right)\right]\right\}\Bigg|_{\substack{p_1 \to i\omega \\ p_2 \to -i\omega/v_x}} \qquad (7)$$

with arbitrary wave component of solution not connected definitely with exciting electrical field, the latter determines roots of the dispersion equation with corresponding them velocities of waves. However, according to the taken condition, the initial (background) Maxwellian distribution function $f_0$ contains no kinematic waves. Besides that, as far as a kinematic wave is not supported by the boundary electric field, some time or other it must be exhausted at anyhow rare collisions. Thus there must be realized asymptotical condition

$$W_{kinem} = 0 \qquad (8)$$

which determines, as it was noted before, the image $f_{p_1}$ of the boundary distribution function $f(\vec{v}, 0, t)$, namely from integral equation

$$v_x f_{p_1} \approx \frac{e v_z f_0}{k_B T} \left( \frac{i v_x E_{p_1}}{\omega} - \frac{v_x^2 F_{p_1}}{\omega^2} + \frac{4\pi e n_e p_1}{c^2} \frac{v_x^2}{\omega^2} \int \frac{v_x v_z f_{p_1}}{p_1 + v_x p_2} d\vec{v} \right)\Bigg|_{p_2 \to -i\omega/v_x}. \tag{9}$$

Let designate in the following

$$f_{p_1} \cdot (p_1 - i\omega) \equiv f_{p_1}^{(0)}; \qquad E_{p_1} \cdot (p_1 - i\omega) = E_0; \qquad F_{p_1} \cdot (p_1 - i\omega) \equiv F_{p_1}^{(0)}. \tag{10}$$

A lucky circumstance here is the possibility to neglect the small integral term of the order $v^2/c^2$ in equation (9). Then the equation for $f_{p_1}^{(0)}$ acquires a simple form

$$v_x f_{p_1}^{(0)} \approx \frac{e}{k_B T} v_z f_0 \cdot \left( \frac{i v_x E_{p_1}^{(0)}}{\omega} - \frac{v_x^2 F_{p_1}^{(0)}}{\omega^2} \right), \tag{11}$$

and the equation for electric field

$$E_{p_1 p_2}^{(0)} G_{p_1 p_2} \approx p_2 E_0 + F_{p_1}^{(0)}, \quad p_1 = i\omega, \tag{12}$$

where

$$E_{p_1 p_2}^{(0)} \equiv E_{p_1 p_2} \cdot (p_1 - i\omega); \quad G_{p_1 p_2} = A_{p_1 p_2} \left( p_2 - p_2^{(1+)} \right)\left( p_2 - p_2^{(1-)} \right)\left( p_2 - p_2^{(2+)} \right)\left( p_2 - p_2^{(2-)} \right), \tag{13}$$

$$A_{p_1 p_2} \equiv \overline{\frac{v_x^2}{\omega^2 + v_x^2 p_2^2}}. \tag{14}$$

Since the fast waves at their propagation leave behind by far the slow waves and in the half-infinite slab problem don't meet any obstacles, there arises a natural assumption that the backward fast waves $\exp\left( i\omega t + i k^{(1+)} x \right)$ are absent, that is (with $p_1 = i\omega$)

$$p_2^{(1+)} E_0 + F_{p_1}^{(0)} = 0, \tag{15}$$

there is correspondingly

$$F_{p_1}^{(0)} \approx -\frac{i\omega}{c} \sqrt{1 - \omega_L^2/\omega^2} E_0. \tag{16}$$

According to residua theorem, amplitude of the fast forward wave $\exp\left( i\omega t + i k^{(1-)} x \right)$ is in this case

$$\frac{1}{A_{p_1 p_2^{(1-)}}} \frac{p_2^{(1-)} E_0 + F_{p_1}^{(0)}}{2 p_2^{(1-)} \left( p_2^{(1)^2} - p_2^{(2)^2} \right)} \approx E_0. \tag{17}$$

In analogous manner for amplitude of the forward slow wave $\exp\left(i\omega t + ik^{(2-)}x\right)$ one obtains

$$\frac{1}{A_{p_1 p_2^{(2-)}}} \frac{p_2^{(2-)}E_0 + F_{p_1}^{(0)}}{2p_2^{(2-)}\left(p_2^{(2)^2} - p_2^{(1)^2}\right)} = \frac{E_0}{4}\frac{\overline{v_x^2}}{c^2}\frac{\omega_L^2}{\omega^2}\left(1 + \frac{\sqrt{\overline{v_x^2}}}{c}\sqrt{1 - \omega_L^2/\omega^2}\right), \tag{18}$$

and for amplitude of the backward slow wave $\exp\left(i\omega t + ik^{(2+)}x\right)$ one obtains

$$\frac{1}{A_{p_1 p_2^{(2+)}}} \frac{p_2^{(2+)}E_0 + F_{p_1}^{(0)}}{2p_2^{(2+)}\left(p_2^{(2)^2} - p_2^{(1)^2}\right)} = \frac{E_0}{4}\frac{\overline{v_x^2}}{c^2}\frac{\omega_L^2}{\omega^2}\left(1 - \frac{\sqrt{\overline{v_x^2}}}{c}\sqrt{1 - \omega_L^2/\omega^2}\right). \tag{19}$$

Defining

$$C \equiv \frac{E_0}{4}\frac{\overline{v_x^2}}{c^2}\frac{\omega_L^2}{\omega^2}\left(1 + \frac{\sqrt{\overline{v_x^2}}}{c}\sqrt{1 - \omega_L^2/\omega^2}\right), \tag{20}$$

$$D \equiv \frac{E_0}{4}\frac{\overline{v_x^2}}{c^2}\frac{\omega_L^2}{\omega^2}\left(1 - \frac{\sqrt{\overline{v_x^2}}}{c}\sqrt{1 - \omega_L^2/\omega^2}\right), \tag{21}$$

one obtains real value solution for the sum of the forward and backward slow waves in the form of identities

$$\begin{aligned}
&C\cos\left(\omega t - kx\right) + D\cos\left(\omega t + kx\right) = C\cos\omega t\cos kx + C\sin\omega t\sin kx + D\cos\omega t\cos kx - \\
&-D\sin\omega t\sin kx = \left(C + D\right)\cos\omega t\cos kx + \left(C - D\right)\sin\omega t\sin kx + \left(C + D\right)\sin\omega t\sin kx - \\
&-\left(C + D\right)\sin\omega t\sin kx = \left(C + D\right)\cos\left(\omega t + kx\right) + 2C\sin\omega t\sin kx = \left(C + D\right)\cos\left(\omega t - kx\right) - \\
&-2D\sin\omega t\sin kx.
\end{aligned} \tag{22}$$

Thus, mathematical multiformity of representing the sum of forward and backward waves as traveling and standing waves might lead to incorrect determination of plasma wave stream and echo effects in dependence on type of an analyzing device. The true characteristics of the real oscillations and streams can be obtained by the simultaneous detection not only traveling waves but also standing waves using corresponding methods and devices.

It is necessary however to note the extremely small amplitude of slow waves, with its ratio to amplitude of fast wave less than $10^{-6}$ at $T = 10000K$.

At that time the ratio of numbers of individual slow and fast waves placed on a given length is

$$\frac{\lambda_{fast}}{\lambda_{slow}} \simeq \frac{c}{\sqrt{\overline{v_x^2}}}\frac{1}{\sqrt{1 - \omega_L^2/\omega^2}}, \tag{23}$$

thus augmenting summed (accumulated) energy of slow waves placed on the same length.

Proceeded consideration leads to the following natural conclusions.

1. The general view for interrelation of fast and slow wave branches and plasma echo can be considerably affected by the presence of reflecting wall at plasma tube exit, especially at differing reflection coefficients for the fast and slow waves. Echo intensities can considerably grow at anyhow combinations of the forward and backward fast waves.

2. Considerable changes can arise in the case of transit regime (in dependence on duration of the first sequence of exciting pulses of electric field and on the moment and distance of measurements of amplitudes of the forward or background, also standing waves).

3. It ought to note the possibility of tangled effects of an echo which arise not by account for, as it is usually accepted (cf. [4]), non-linear interactions of waves or accidental matching phases of kinematic waves, represented as modulated beams of resonance electrons (Van Kampen waves), but in the frame of just simple linear approach. So, these effects might be connected with athe generation and decay of standing and traveling non interacting waves with amplitudes set by the boundary field, possibly with releasing energy accumulated in standing waves at the boundary of plasma slab after shutting down exciting field.

For instance, at having settled standing waves of the slow branch with $\lambda_{slow} = 2\pi \sqrt{\overline{v_x^2}} \big/ \omega,$ after shutting down the field $E_0 \cos \omega t$ and "running away" fast wave, the solution of now initial problem is defined by the same dispersion expression (4) with $G_{p_1 p_2} = 0.$

This equation in the identical form is

$$\frac{p_1^4 - p_1^2 \left( c^2 p_2^2 + \overline{v_x^2} p_2^2 - \omega_L^2 \right) + c^2 p_2^4 \overline{v_x^2}}{c^2 \left( p_1^2 - \overline{v_x^2} p_2^2 \right)} = 0, \tag{24}$$

and after substituting the root $p_2^{(2\pm)} = \pm i\omega \big/ \sqrt{\overline{v_x^2}}$ has besides the roots $p_1^{(1\pm)} = \pm i\omega$ also approximate roots

$$p_1^{(2\pm)} \simeq \pm i\omega c \big/ \sqrt{\overline{v_x^2}}, \tag{25}$$

what means existing of a new high frequency branch with the frequency $\omega_1 = \omega c \big/ \sqrt{\overline{v_x^2}}$ and the phase and group velocities equal $c$.

The possibility of non linear wave interactions with hybridization of both branches of slow and fast electron waves was noted in [5].

An attempt to analyzing pulse exciting signal and its backward echo response is proposed in [6].